\documentstyle{article}

\include{epsf}

\begin{document}

\title{Statistical Derivation of the Evolution Equation \\ of Liquid Water Path
Fluctuations in Clouds}

\author{K. Ivanova$^{1}$ and M. Ausloos$^{2}$ \\ \\ 
$^{1}$ Department of
Meteorology, Pennsylvania State University, \\ University Park, PA 16802, USA\\
$^{2}$ SUPRAS and GRASP, B5, Sart Tilman Campus,\\ B-4000 Li$\grave e$ge, Belgium}



\maketitle


\begin{abstract} How to distinguish and quantify deterministic and random
influences on the statistics of turbulence data in meteorology cases is discussed
from first principles. Liquid water path (LWP) changes in clouds, as retrieved
from radio signals, upon different delay times, can be regarded as a stochastic
Markov process. A detrended fluctuation analysis method indicates the existence
of long range time correlations. The Fokker-Planck equation which models very
precisely the LWP $fluctuation$ empirical probability distributions, in
particular, their non-Gaussian heavy tails is explicitly derived and written in
terms of a drift and a diffusion coefficient. Furthermore, Kramers-Moyal
coefficients, as estimated from the empirical data, are found to be in good
agreement with their first principle derivation. Finally, the equivalent Langevin
equation is written for the LWP increments themselves. Thus rather than the
existence of hierarchical structures, like an energy cascade process, {\it strong
correlations} on different $time$ $scales$, from small to large ones, are
considered to be proven as intrinsic ingredients of such cloud evolutions.

\end{abstract}

\vskip 0.5cm


\vskip 0.5cm

\section{Introduction}

In order to establish a sound understanding for any scientific phenomenon, one
has from numerical data to obtain laws which can be next derived theoretically.
Some difficulty arises in particular in nonlinear dynamical systems because of
the problem to sort out noise from both chaos and deterministic components
[Bhattacharya and Kanjilal, 2000; Provenzale et al., 1992]. Moreover, algorithms
for solving the inverse problem for nonlinear systems with underlying unknown
dynamical processes are notoriously hardly reliable [Theiler, 1992]. In fact, to
extract dynamical equations for chaotic-like data is an enormous challenge
[Rowlands, 1992]. Practically one is often led to empirical relationships. This
is the case of the meteorology field where there is a widely mixed set of various
(sometimes) unknown influences, over different time and space scales. Works of
interest for sorting out deterministic ingredients in chaotic systems, as in
[Hilborn, 1994; Ott, 1993; Schreiber, 1999; Celluci, 1997; Davis et al., 1996a],
can be mentioned for general purpose though this is not an exhaustive list of
references.

It is known that two equivalent $master$ $equations$ govern the dynamics of a
system, i.e. the Fokker-Planck equation and the Langevin equation, the former for
the probability distribution function of time and space signal increments, the
latter for the increment themselves [Reichl, 1980; Ernst et al., 1969; Risken,
1984; H\"aggi and Thomas, 1982; Gardiner, 1983]. They are both condensates of the
huge set of (6N) Hamilton equations which should in practice describe the whole
dynamics of the system of N particles by giving the time evolution of both
coordinates and momenta of each individual particle. This is a highly unrealistic
scheme of work, and therefore conservation laws are used in order to derive the
Navier-Stokes equations, from the averaging of basic quantities weighted by the
above probability functions [Reichl, 1980; Huang, 1967]. Such a
"micro-turned-macro-scopic description" can be by passed when describing the {\it
evolution of the quasi-equilibrium probability distribution function} of such
particles, i.e. in writing a Boltzmann equation [Huang, 1967]. However this
corresponds to a mean field description of the Fokker-Planck equation (FPE)
[Reichl, 1980; Ernst et al., 1969; Risken, 1984; H\"aggi and Thomas, 1982;
Gardiner, 1983], an approximation which might be too rough for this sort of
nonlinear system dynamics.

Recently much advance has been made in describing nonlinear phenomena in
meteorology, after the canonical Lorenz approximation [Lorenz, 1963; Schuster,
1984] of the Navier-Stokes equations for sorting out some deterministic and
stochastic components of turbulent fluid motion in the atmosphere. In fact,
turbulence [Frisch, 1995; Parisi and Frisch, 1985; Hunt, 1999] seems now to be
part of more general class of problems similar to those found when there are long
range fluctuations, as in financial data [Ghashghaie et al., 1996; Mantegna and
Stanley, 1995], traffic [Chowdhury et al., 2000], dielectric breakdown
[Vandewalle et al., 1999], heart beats [Ivanov et al., 1999], etc... Such
observations corroborate the idea that scaling [Stanley, 1971] and fractal
geometry [Schuster, 1984] principles could be useful in such studies on
atmospheric turbulence [Marshak et al., 1997]. Thus to derive the FPE [Risken,
1984; H\"aggi and Thomas, 1982; Gardiner, 1983] for some meteorological effect is
of primary interest.

The aim of this paper is to derive such an FPE, in a first principle statistical approach, from given raw data, i.e. for the
Liquid Water Path (LWP) fluctuations in clouds, - in terms of a drift $D^{(1)}$
and a diffusion $D^{(2)}$ coefficient. Much following the lines of thought of [Friedrich et al., 2000] it will be shown that the FPE (also known as
the Kolmogorov equation) can be derived up to the {\it first two} moments of the
({\it measured}) conditional probability distribution functions $p(\Delta
x,\Delta t)$, where $\Delta x$ is a signal increment and $\Delta t$ a time delay.
Thus the method is not anymore based on the conventional phenomenological
comparison between models and several stochastic aspects, but is a $model$
$independent$ (or first principle, {\it from a statistical point of view})
approach. This allows us to examine long and short time scales on the {\it same
footing}, and suggests new features to be implemented in related weather
forecasting. Indeed {\it it is here below verified} that the solution of the FPE
yields the probability distributions with high accuracy, including the long and
high tail $\Delta x$ and $\Delta t$ events. In some sense the agreement proves that searching for more improvement with including further event of similar cases is not necessary.

Furthermore the so found analytical form of both drift $D^{(1)}$ and diffusion
$D^{(2)}$ coefficients has a simple physical interpretation. It is also shown
that a truncation of the FPE expansion in terms of high order joint probabilities
to the first two terms is quite sufficient. Moreover the identification of the
underlying process leading to the heavy tailed probability density functions
(pdf) for the LWP fluctuation correlation with $\Delta t$ and the volatility
clustering (as seen in Fig. 3 below) seems to be a new observation leading to an
interestingly new set of puzzles for this sort of clouds [Cahalan et al., 1995],
and more generally in atmospheric science.

\section{Data and theoretical analysis}

The changes in a time series for a signal $x(t)$ are commonly measured by the
quantity $r = x(t + \Delta t)/x(t)$, or increments $\Delta x$ = $x(t + \Delta t)
-x(t)$. Results of the analysis of a data set $r(t)$, selecting among others LWP
observations, are presented here. The 25 772 data points are taken from microwave
radiometer measurements at the Southern Great Plains site of Atmospheric
Radiation Measurement (ARM)\footnote{$http://www.arm.gov$} of Department of
Energy during the period January 9-14, 1998 as used by [Ivanova et al., 2000].
The microwave radiometer measures the brightness temperature at two frequency
channels, one at 23.8~GHz and the other at 31.4~GHz. Then both brightness
temperature data series are used to retrieve the vertical columnar amount of
water vapor and vertical columnar amount of liquid water in the cloud, i.e. the
so-called liquid water path. The raw data $r_{i}$ ($i$ is equivalent to a time
index) of the liquid water path in stratus clouds is shown in Fig. 1.

The first thing is to consider whether the distribution function of increments
has short or long tails, and whether there is some scaling law involved. Next the
central issue is the understanding of the statistics of fluctuations which
determine the evolution of the cloud.

There are different estimators for the short and/or long range dependence of
fluctuations correlations [Taqqu et al., 1995]. Through a detrended fluctuation
analysis (DFA) method [Ausloos and Ivanova, 1999] we show first that the {\it
long range correlations} are not brownian. The method has been used previously in
the meteorological field [Ivanova et al., 2000; Ausloos and Ivanova, 1999;
Koscielny-Bunde et al., 1998; Ivanova and Ausloos, 1999] and its concepts are not
repeated here. We show on Fig. 2a the final result, for the

\begin{equation} <F^2(\tau)>^{1/2} = \sqrt{{1 \over \tau }
{\sum_{n=k\tau+1}^{(k+1)\tau} {\left[r(n)- z(n)\right]}^2}} \sim \tau^{\alpha}
\end{equation}

\noindent indicating a scaling law characterized by an exponent
$\alpha=0.36\pm0.01$, markedly different from $0.5$, thus indicating
antipersistence of the signal from about 3 to about 150~minutes. Next the power
spectrum $S(f)\sim f^{-\beta}$ with spectral exponent {\bf $\beta=1.57\pm 0.02$}
is also shown (see Fig. 2b) [Davis et al., 1996b]. A Kolmogorov-Smirnov test on
surrogate data has indicated the statistical validity of the value and error
bars.

Thereafter we focus on the LWP changes measured as increments $\Delta x$ which
are given in Fig. 3 in units of the standard deviation $\sigma$ of $\Delta x$ at
$\Delta t=640$~s. ln order to characterize the statistics of these LWP changes,
LWP increments $\Delta x_1$, $\Delta x_2$ for delay times $\Delta t_1$, $\Delta
t_2$ at the same time $t$ are considered. The corresponding joint probability
density functions are evaluated for various time delays $\Delta t_1$ $<$ $\Delta
t_2$ $<$ $\Delta t_3$ $<$ ... directly from the given data set. One example of a
contour plot of these functions is exhibited in Fig. 4. If two LWP changes, i.e.
$\Delta x_1$ and $\Delta x_2$ are statistically independent, the joint pdf should
factorize into a product of two probability density functions:

\begin{equation} p ( \Delta x_1 , \Delta t_1 ; \Delta x_2 , \Delta t_2 ) = p (
\Delta x_1 , \Delta t_1 ) p(\Delta x_2 , \Delta t_2 ). \end{equation}

\noindent leading to a camelback or an isotropic single hill landscape. The
tilted anisotropic form of the joint probability density (Fig. 4) clearly shows
that such a factorization does not hold for small values of $|$log($\Delta t_1/
\Delta t_2$)$|$, whence both LWP changes are {\it statistically dependent}. The
same is found to be true for other ($\Delta t_i/ \Delta t_j$) ratios. This is in
agreement with the hint taken from the above observations for the long range
cross-correlation functions with the DFA.

This implies a (fractal-like) hierarchy of $time$ scales. If fluctuations in LWP
went up over a certain $\Delta t_2$, then it is more likely that, on a shorter
$\Delta t_1$ within the larger one, the LWP went down instead of up. To analyze
these correlations in more detail, the question on what kind of statistical
process underlies the LWP changes over a $series$ of nested time delays $\Delta
t_i$ of decreasing duration should be raised. A complete characterization of the
statistical properties of the data set in general requires the evaluation of
joint pdf's $p^N$($\Delta x_1$,$\Delta t_1$;...;$\Delta x_N$,$\Delta t_N$)
depending on N variables (for arbitrarily large N). In the case of a Markov
process (a process without memory) [Schuster, 1984], an important simplification
arises: The N-point pdf $p^N$ is generated by a product of the conditional
probabilities $p$($\Delta x_{i+l}$,$\Delta t_{i+l}|\Delta x_{i}$,$\Delta t_{i}$)
= $p$($\Delta x_{i+l}$,$\Delta t_{i+l}$;$\Delta x_{i}$,$\Delta
t_{i}$)/ $p$($\Delta x_{i}$,$\Delta t_{i}$) for i = 1,...,N-1. The conditional
probability is given by the probability of finding $\Delta x_{i+1}$ values for
fixed $\Delta x_{i}$. As a necessary condition, the Chapman-Kolmogorov equation

\begin{equation} p(\Delta x_{1},\Delta t_{1} | \Delta x_{2},\Delta t_{2}) = \int
d(\Delta x_{i}) p(\Delta x_{1},\Delta t_{1} | \Delta x_{i},\Delta t_{i}) p(\Delta
x_{i},\Delta t_{i} | \Delta x_{2},\Delta t_{2}) \end{equation}

\noindent should hold for any value of $\Delta t_i$, with $\Delta t_1$ $<$
$\Delta t_i$ $<$ $\Delta t_2$. We checked the validity of the Chapman-Kolmogorov
equation for different $\Delta t_{i}$ triplets by comparing the directly
evaluated conditional probability distributions $p$($\Delta x_1,\Delta t_1|\Delta
x_2,\Delta t_2$) with the ones calculated ($P_{cal}$) according to Eq.(3), in
Fig.5. The solutions of Eq. (4) usually give rise to exponential laws, but
initial conditions must be taken into account for adjusting the constants of the
integration. Starting with a stretched exponential, the latter is conserved after
successive integrations, with a width  varying with $\Delta t$.   The power law
probability density function {\it which is observed} is thus a signature of
fractal properties of the signal increments because the tails of the pdf are not
truly exponential ones. In Fig. 5a, the contour lines of the two
corresponding pdf's for all values of $\Delta x_1$ are shown. There are mild
deviations, probably resulting from a finite resolution of the statistics. Cuts
for some exemplarily chosen values of $\Delta x_1$ are shown in addition in Fig.
5 (b-d).

As is well known, the Chapman-Kolmogorov equation yields an evolution equation
for the change of the distribution functions $p$($\Delta x,\Delta t|\Delta
x_1,\Delta t_1$) and $p$($\Delta x$,$\Delta t$) across the scales $\Delta t$
[Reichl, 1980; Ernst et al., 1969; Risken, 1984; H\"aggi and Thomas, 1982;
Gardiner, 1983]. For the following it is convenient (and without loss of
generality) to consider a normalized logarithmic time scale $\tau$= $ln$
(640/$\Delta t$). The limiting case $\Delta t_{i}$ $\rightarrow$ 0 corresponds to
$\tau$ $\rightarrow$ $\infty$.

The Chapman-Kolmogorov equation formulated in differential form yields a master
equation, which can take the form of a Fokker-Planck equation (for a detailed
discussion, we refer the reader to [Reichl, 1980; Ernst et al., 1969; Risken,
1984; H\"aggi and Thomas, 1982; Gardiner, 1983])\footnote{The change in variable
from $\Delta t$ to $\tau$ is non linear and might be debatable if the following
equations (4) and (9) are integrated with respect to time. This is not done here.
The integrations discussed in the text are made at fixed $\Delta t$ or $\tau$.}:

\begin{equation} \frac {d}{d\tau} p (\Delta x, \tau)= [ -
\frac{\partial}{\partial \Delta x } D^{(1)} ( \Delta x , \tau ) +
\frac{\partial}{\partial ^2 \Delta x^2} D^{(2)}(\Delta x,\tau)] p (\Delta x,
\tau) \end{equation}

\noindent in terms of a drift $D^{(1)}$($\Delta x$,$\tau$) and a diffusion
coefficient $D^{(2)}$($\Delta x$,$\tau$). Note that the Fokker-P1anck equation
results from a truncation of the master equation expanded in  terms of high order
joint probabilities. According to Pawula's theorem [Risken, 1984],  such a
truncation is valid provided that the fourth order coefficient $D^{(4)}$
vanishes. We did check  the $D^{(4)}$ coefficient for $\Delta x \in [-0.3,0.3]$
and obtained values that are two orders of magnitude smaller than $D^{(2)}$ and
three decades smaller than $D^{(1)}$, which we consider justifies the truncation.
The functional dependence of the drift and diffusion coefficients can be 
estimated directly from the moments $M^{(k)}$ of the conditional 
probability distributions (cf. Fig. 5):

\begin{equation} M^{(k)} = \frac {1} {\Delta \tau} \int d\Delta x^{'} (\Delta
x^{'} - \Delta x)^{k} p(\Delta x^{'},\tau + \Delta \tau | \Delta x,\tau)
\end{equation}

\noindent for different small $\Delta \tau$'s ( Fig. 6), such that

\begin{equation} D^{(k)} (\Delta x,\tau) = \frac {1}{k!} \mbox{lim} M^{(k)}
\end{equation}

\noindent for $\Delta \tau \rightarrow 0$.

The coefficient $D^{(1)}$ shows a linear dependence on $\Delta x$, while
$D^{(2)}$ can be approximated by a polynomial of degree two in $\Delta x$. This
behavior was found for all scales $\tau$ and $\Delta \tau$. Therefore the drift
term $D^{(1)}$ is well approximated by a linear function of $\Delta x$, whereas
the diffusion term $D^{(2)}$ follows a function quadratic in $\Delta x$. For
large values of $\Delta x$ the statistics becomes poorer and the uncertainty
increases. From a careful analysis of the data based on the functional
dependences of $D^{(1)}$ and $D^{(2)}$ (Fig. 6 a-b), the following approximations
hold true:

\begin{equation} D^{(1)} = - 0.0078 \Delta x, \end{equation}

\begin{equation} D^{(2)} = 0.00259( \Delta x - 0.0305)^2 +0.00009. \end{equation}

It may be worthwhile to remark that the observed quadratic dependence of the
diffusion term $D^{(2)}$ is essential for the logarithmic scaling of the
intermittency parameter in previous studies on turbulence. Finally, the FPE for
the distribution function is known to be equivalent to a Langevin equation for
the variable, i.e. $\Delta x$ here, (within the Ito interpretation [Reichl, 1980;
Ernst et al., 1969; Risken, 1984; H\"aggi and Thomas, 1982; Gardiner, 1983])

\begin{equation} \frac {d}{d\tau} \Delta x(\tau) = D^{(1)}(\Delta x(\tau),\tau) +
\eta(\tau) \sqrt {{D^{(2)} (\Delta x(\tau),\tau)}}, \end{equation}

\noindent where $\eta(\tau)$ is a fluctuating $\delta$-correlated force with
Gaussian statistics, i.e. $<$ $\eta(\tau)$ $\eta(\tau')$$>$ = 2$ \delta (\tau
-\tau')$.

\section{Conclusions}

At least since the pioneering work of Lorenz [Lorenz, 1963] stochastic problems
in turbulence are commonly treated as processes running in time $t$ with long
time correlations. Inspired by the idea of an existing energy cascade process
[Cahalan, 1994] we present here a new approach, namely, we investigate among
other related phenomena, how LWP changes are correlated on different time steps
$\Delta t$. The pdf shape expresses an unexpected high probability (compared to a
Gaussian pdf) of large LWP changes which is of utmost importance for forecasting.
In recent works [Friedrich et al., 2000; Friedrich and Peinke, 1997] this finding
leads to postulating the existence of hierarchical features, i.e. an energy
cascade process from large to small time scales. The existence of finite non
Gausssian tails for large events is thought to be due or to imply drastic
evolutions, as for earthquakes, financial crashes or heart attacks.

One unexpected result is the time dependence of the tails (Fig.3) which seems
quite smooth for the LWP case. This puzzle should initiate new research with a
goal toward forecasting, in particular to find cases in which such a smoothness
does not hold. This evolution shows {\it how the pdf's deviate more and more from
a Gaussian shape} as $\Delta t$ increases or $\tau$ decreases. This definitely is
a new quality in describing the hierarchical structure of such data sets, - not
seen in the DFA nor spectral result. Now it becomes clear that one must not
require stationary probability distributions for LWP differences for different
$\tau$; on the contrary the coupling between different scales $\tau$ via a Markov
process is essential. Thus also a proper modeling of the time evolution of LWP
differences for a fixed time delay must take into account the coupling of these
quantities to LWP differences on different time delays.

Furthermore, it is stressed that in contrast to the use of phenomenological
fitting functions, the above method provides the evolution process of pdf's {\it
from small time delays to larger ones}. Interestingly this is through an analogy
with two physically meaningful coefficients, a drift term $D^{(1)}$ and a
diffusion term $D^{(2)}$. The first one linearly behaves, thus looks like a
"restoring force", the second behaving quadratically in $\Delta x$, is obviously
like an autocorrelation function as for {\it bona fide} (chemical) diffusion. Of
course further theoretical work is needed before understanding the numerical
values in Eqs. (7)-(8).

Finally, the present report presents a method on how to derive an underlying
mathematical ({\it statistical or model free}) equation for a LWP cascade
directly. The method yields an effective stochastic equation in the form of a FPE
in the variable $\Delta t$. The excellent agreement between the experimental and
the statistical approach removes the need for much further proof based on
examining many other cases for the same type of clouds.  The FPE provides the
complete knowledge as to how the statistics of LWP distribution change
correlations on different delay times. Since this includes an autocorrelation
analysis in time $t$ for a scalar $\Delta x$, it is suggesting that the findings
could be implemented in atmospheric weather low dimensional $vector-models$
[Grabowski and Smolarkiewicz, 1999; Ragwitz and Kantz, 2000]. Consequently, even
though for the first time in atmospheric science, a stochastic equation (of LWP
evolution) is hereby introduced from first principles, several interesting
statements can be presented and new questions opened.

\vskip 1.6cm {\bf Acknowledgments} \vskip 0.6cm

We thank C. Nicolis, T.P. Ackerman, H.N. Shirer and E.E. Clothiaux for
stimulating discussions and comments. Email correspondence with J. Peinke, Ch.
Renner and R. Friedrich is specifically appreciated. This investigation was
partially supported by Battelle grant number 327421-A-N4.

\vskip 1cm

\begin{center} {\bf REFERENCES} \end{center}

\vspace*{0.4cm}

Ausloos, M., and K. Ivanova, 1999: Precise (m,k)-Zipf diagram analysis of
mathematical and financial time series when $m = 6$, $k = 2$, {\it Physica A},
{\bf 270}, 526-543.

\vspace*{0.4cm}

Bhattacharya, J., and P.P. Kanjilal, 2000: Revisiting the role of correlation
coefficient to distinguish chaos from noise {\it Eur. Phys. J. B}, {\bf 13},
399-403.

\vspace*{0.4cm}

Cahalan, R.F., 1994: Bounded cascade clouds: albedo and effective thickness, {\it
Nonlinear Proc. Geophys.}, {\bf 1}, 156-167.

\vspace*{0.4cm}

Cahalan, R.F., D. Silberstein and J. B. Snider, 1995: Liquid water path and
plane-parallel albedo bias during ASTEX, {\it J. Atmos. Sci.}, {\bf 52},
3002-3012.

\vspace*{0.4cm}

Cellucci, C.J., A.M. Albano, P.E. Rapp, R.A. Pittenger, and R.C. Josiassen, 1997:
Detecting noise in a time series, {\it Chaos}, {\bf 7}, 414-422.

\vspace*{0.4cm}

Chowdhury, D., L. Santen, and A. Schadschneider, 2000: Statistical physics of
vehicular traffic and some related systems, {\it Phys. Rep.}, {\bf 329}, 199-329.

\vspace*{0.4cm}

Davis, A., A. Marshak, W. J. Wiscombe, and R. F. Cahalan, 1996a: Multifractal
Characterizations of Intermittency in Nonstationary Geophysical Signals and
Fields - A Model-Based Perspective on Ergodicity Issues Illustrated with Cloud
Data in {\it Current Topics in Nonstationary Analysis}, Eds. G. Trevino, J.
Hardin, B. Douglas, and E. Andreas, (World Scientific, Singapore, 1996) 97-158.

\vspace*{0.4cm}

Davis, A., A. Marshak, W. Wiscombe and R. Cahalan, 1996b: Scale-invariance of
liquid water distributions in marine stratocumulus. Part I: Spectral properties
and stationarity issues, {\it J. Atmos. Sci.}, {\bf 53}, 1538-1558.

\vspace*{0.4cm}

Ernst, M.H., L.K. Haines, and J.R. Dorfman, 1969: Theory of transport
coefficients for moderately dense gases, {\it Rev. Mod. Phys.}, {\bf 41},
296-316.

\vspace*{0.4cm}

Friedrich, R., and J. Peinke, 1997:Description of a Turbulent Cascade by a
Fokker-Planck Equation {\it Phys. Rev. Lett.} {\bf 78}, 863-866.

\vspace*{0.4cm}

Friedrich, R., J. Peinke, and Ch. Renner, 2000: How to quantify deterministic and
random influences on the statistics of the foreign exchange market, {\it Phys.
Rev. Lett.}, {\bf 84}, 5224-5228.

\vspace*{0.4cm}

Frisch, U., 1995: {\it Turbulence} Cambridge Univ. Press, Cambridge UK.

\vspace*{0.4cm}

Gardiner, C.W., 1983: {\it Handbook of Stochastic Methods}, Springer-Verlag,
Berlin.

\vspace*{0.4cm}

Ghashghaie, S., W. Breymann, J. Peinke, P. Talkner and Y. Dodge, 1996: Foreign
Exchange Market - A Turbulent Process?, {\it Nature}, {\bf 381}, 767-770.

\vspace*{0.4cm}

Grabowski, W.W., and P.K. Smolarkiewicz, 1999: CRCP: a cloud resolving convection
parameterization for modeling the tropical convecting atmosphere, {\it Physica
D}, {\bf 133}, 171-178.

\vspace*{0.4cm}

H\"anggi, P., and H. Thomas, 1982: Stochastic processes : time evolution,
symmetries and linear response, {\it Phys. Rep.}, {\bf 88}, 207-319.

\vspace*{0.4cm}

Hilborn, R. C., 1994: {\it Chaos and Nonlinear Dynamics}, Oxford University
Press, New York

\vspace*{0.4cm}

Huang, K., 1967: {\it Statistical Mechanics} J. Wiley, New York.

\vspace*{0.4cm}

Hunt, J.C.R., 1999: Environmental forecasting and turbulence modeling {\it
Physica D}, {\bf 133}, 270-295.

\vspace*{0.4cm}

Ivanov, P.Ch., L.A.N. Amaral, A.L. Goldberger, S. Havlin, M.G. Rosenblum, Z.
Struzik, H.E. Stanley, 1999: Multifractality in human heartbeat dynamics{\it
Nature}, {\bf 399}, 461-465.

\vspace*{0.4cm}

Ivanova, K., and M. Ausloos, 1999: Application of the Detrended Fluctuation
Analysis (DFA) method for describing cloud breaking, in {\it Applications of
Statistical Mechanics}, Proc. of a NATO ARW Budapest 1999, A. Gadomski, J.
Kertesz, H.E. Stanley, and N. Vandewalle, Eds., {\it Physica A }, {\bf 274},
349-354.

\vspace*{0.4cm}

Ivanova, K., M. Ausloos, E.E. Clothiaux, and T.P. Ackerman, 2000: Break-up of
stratus cloud structure predicted from non-Brownian motion liquid water and
brightness temperature fluctuations, {\it Europhys. Lett.,} {\bf 52} (1), 40-46.

\vspace*{0.4cm}

Koscielny-Bunde, E., A. Bunde, S. Havlin, H. E. Roman, Y. Goldreich, and H.-J.
Schellnhuber, 1998: Indication of a Universal Persistence Law Governing
Atmospheric Variability {\it Phys. Rev. Lett.}, {\bf 81}, 729-732.

\vspace*{0.4cm}

Lorenz, E.N., 1963: Deterministic nonperiodic flow {\it J. Atmos. Sci.}, {\bf
20}, 130.

\vspace*{0.4cm}

Mantegna, R.N., and H.E. Stanley, 1995: Scaling behavior in the dynamics of an
economic index, {\it Nature}, {\bf 376}, 46-49.

\vspace*{0.4cm}

Marshak, A., A. Davis, W. J. Wiscombe, and R. F. Cahalan, 1997: Scale-invariance
of liquid water distributions in marine stratocumulus. Part II: Multifractal
properties and intermittency issues, {\it J. Atmos. Sci.}, {\bf 54}, 1423-1444.

\vspace*{0.4cm}

Ott, E., 1993: {\it Chaos in Dynamical Systems}, Cambridge University Press

\vspace*{0.4cm}

Parisi, G., and U. Frisch, 1985: in {\it Turbulence and Predictability in
Geophysical Fluid Dynamics and Climate Dynamics}, Ghil M., Benzi R. and Parisi
G., Eds. North Holland, New York.

\vspace*{0.4cm}

Provenzale, A., L.A. Smith, R. Vio, and G. Murante, 1992: Distinguishing between
low-dimensional dynamics and randomness in measured time series, {\it Physica D
}, {\bf 58}, 31-49.

\vspace*{0.4cm}

Ragwitz, M., and H. Kantz, 2000: Detecting non-linear structure and predicting
turbulent gusts in surface wind velocities, {\it Europhys. Lett.}, {\bf 51},
595-601.

\vspace*{0.4cm}

Reichl, L.E., 1980: {\it A Modern Course in Statistical Physics} Univ. Texas
Press, Austin.

\vspace*{0.4cm}

Risken, H., 1984: {\it The Fokker-Planck Equation} Springer-Verlag, Berlin.

\vspace*{0.4cm}

Rowlands, G., and J.C. Sprott, 1992: Extraction of dynamical equations from
chaotic data {\it Physica D}, {\bf 58}, 251-259.

\vspace*{0.4cm}

Schreiber, Th., 1999: Interdisciplinary application of nonlinear time series
methods {\it Phys. Rep.}, {\bf 308}, 1-64.

\vspace*{0.4cm}

Schuster, H.G., 1984: {\it Deterministic Chaos} Physik-Verlag, Weinheim.

\vspace*{0.4cm}

Stanley, H.E., 1971: {\it Phase transitions and critical phenomena} Oxford Univ.
Press, Oxford.

\vspace*{0.4cm}

Taqqu, M.S., V. Teverovsky, and W. Willinger, 1995: Estimators for long-range
dependence: an empirical study, {\it Fractals}, {\bf 3}, 785-798.

\vspace*{0.4cm}

Theiler, J., S. Eubank, A. Longtin, B. Galdrikian and J. D. Farmer, 1992: Testing
for nonlinearity in time series: the method of surrogate data, {\it Physica D},
{\bf 58}, 77-94.

\vspace*{0.4cm}

Vandewalle, N., M. Ausloos, M. Houssa, P.W. Mertens and M.M. Heyns, 1999:
Non-Gaussian behavior and anticorrelations in ultra-thin gate oxides after soft
Breakdown{\it Appl. Phys. Lett.}, {\bf 74}, 1579-1581.

\newpage

\begin{figure}[htb]
\begin{center}
\leavevmode
\epsfysize=8cm
\epsffile{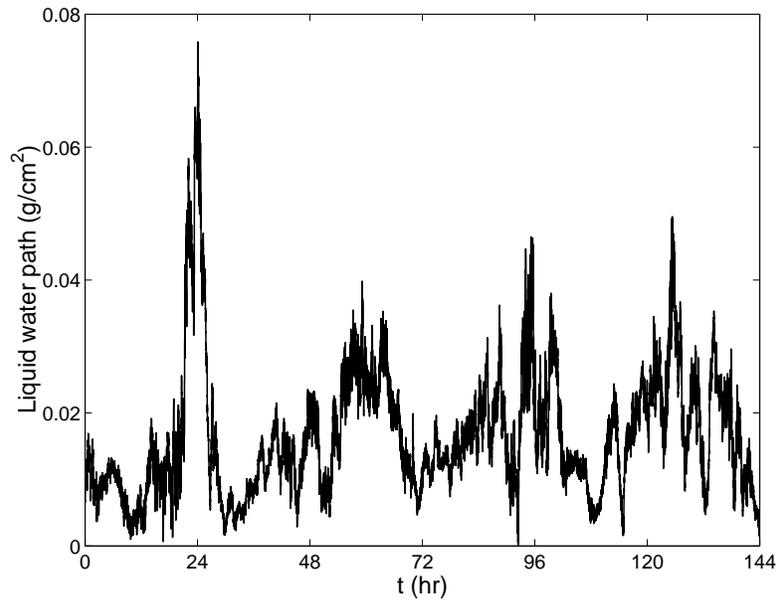}
\end{center}
\caption{Time dependence of a stratus cloud liquid water path
(LWP) from Jan. 9 to 14, 1998, obtained at the ARM Southern Great Plains site
with time resolution of 20~s.}
\end{figure}

\begin{figure}[htb]
\begin{center}
\leavevmode
\epsfysize=8cm
\epsffile{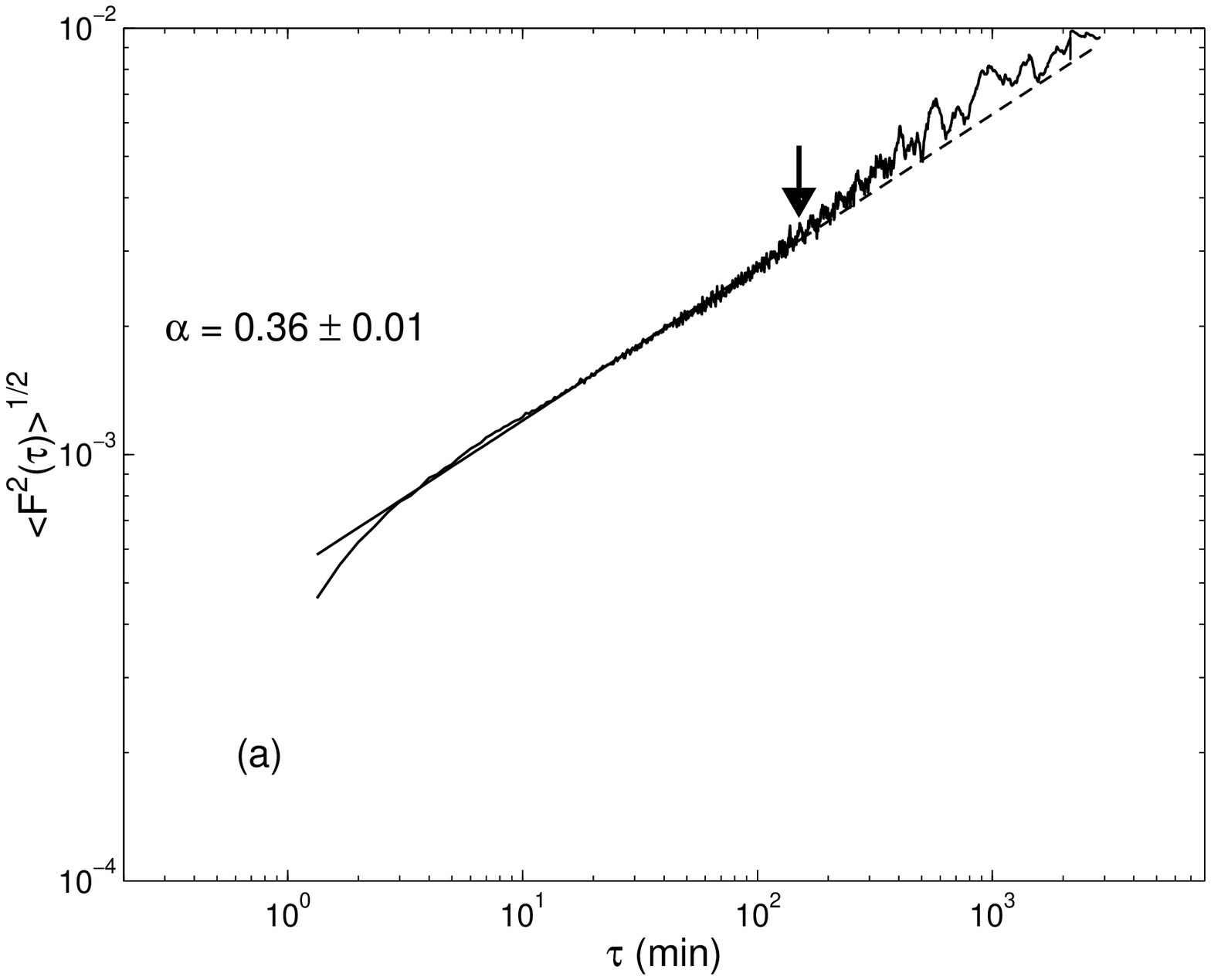}
\vfill
\leavevmode
\epsfysize=8cm
\epsffile{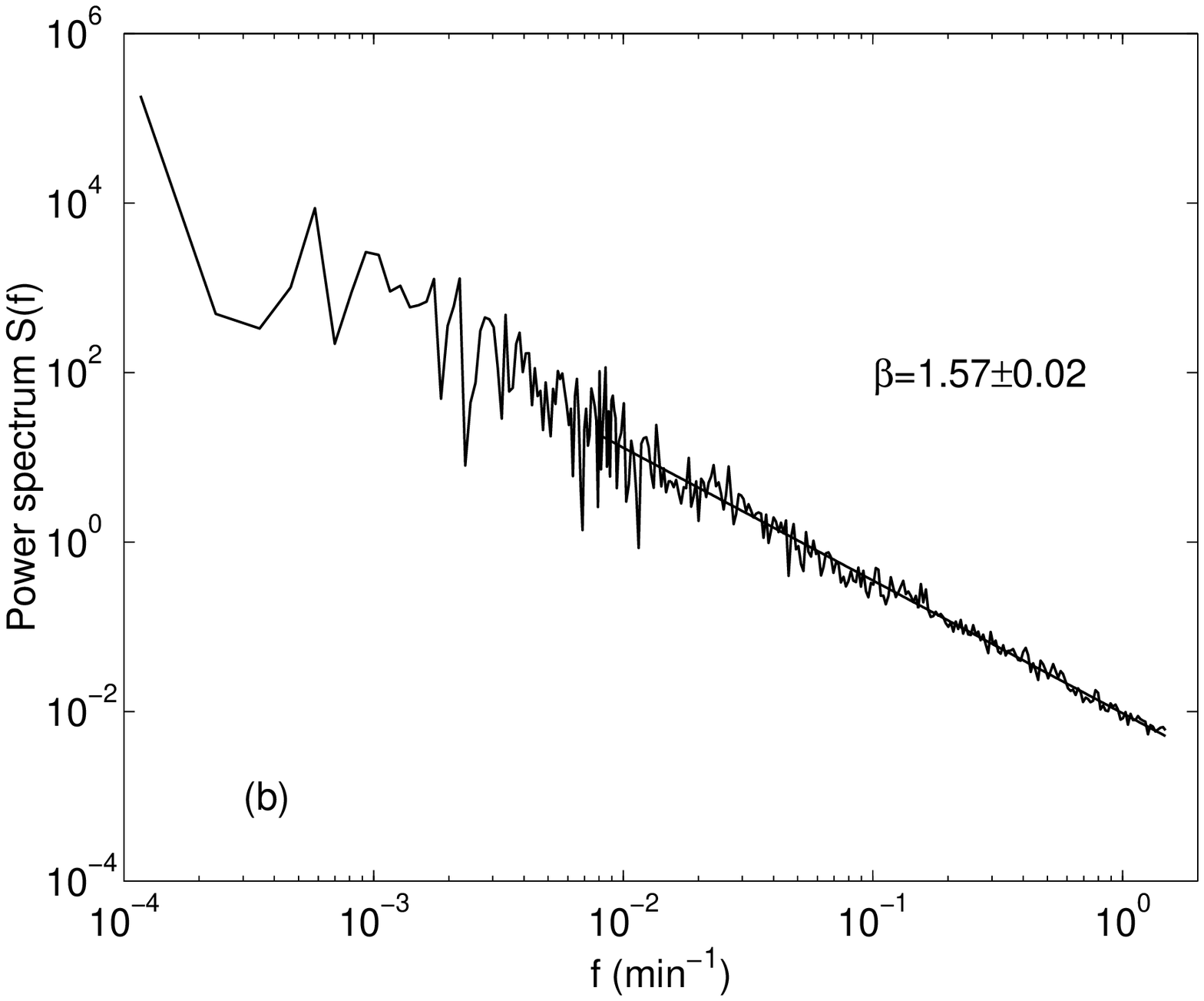}
\end{center}
\caption{(a) The detrended fluctuation analysis
(DFA) function $<F^2(\tau)>^{1/2}$ (Eq. (1)) for the data in Fig. 1. Scaling
range holds for a time interval between ca. 3 min and ca. 150 min. (b) Power
spectrum $S(f)$ for the data in Fig. 1.}
\end{figure}

\begin{figure}[htb]
\begin{center}
\leavevmode
\epsfysize=8cm
\epsffile{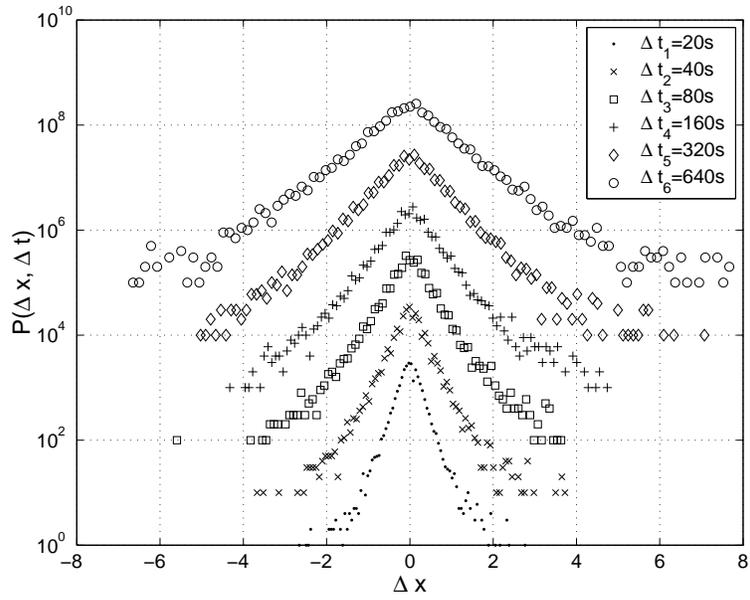}
\end{center} 
\caption{Frequency $p(\Delta x,\Delta t)$ of
the LWP increments $\Delta x$ for different time delays $\Delta t$; the units of
$\Delta x$ are taken as multiples of the standard deviation $\sigma=0.0036$ at
$\Delta t=640$~s.}
\end{figure}

\begin{figure}[htb]
\begin{center}
\leavevmode
\epsfysize=8cm
\epsffile{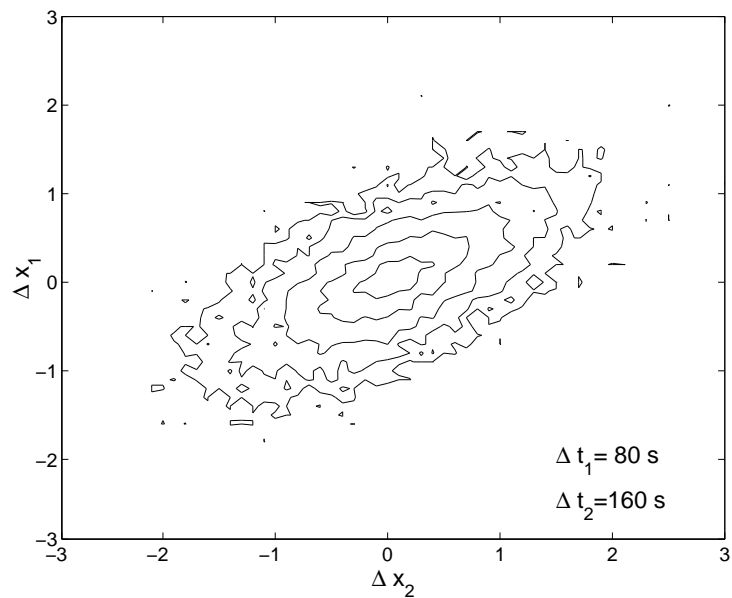}
\end{center} 
\caption{Contour plot of the joint LWP increment
pdf $p(\Delta x_1,\Delta t_1;\Delta x_2,\Delta t_2)$ for the simultaneous
occurrence of LWP fluctuations $\Delta x_1(\Delta t_1)$ and $\Delta x_2(\Delta
t_2)$; $\Delta t_1=80$~s and $\Delta t_2=160$~s. The contour lines correspond to
$log_{10}p=-2,-2.5, -3, -3.5, -4$.}
\end{figure}

\begin{figure}[ht]
\begin{center}
\leavevmode
\epsfysize=8cm
\epsffile{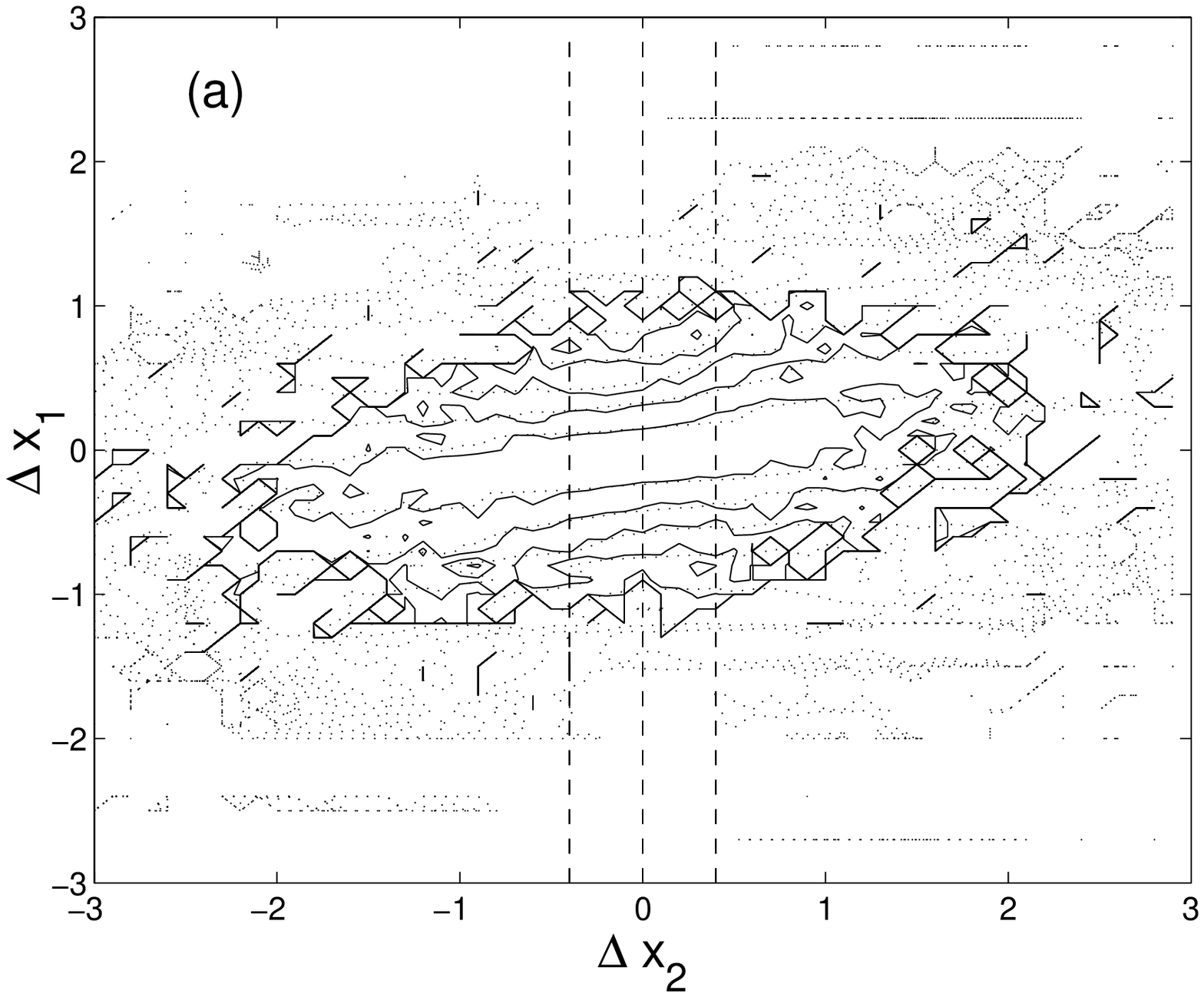}
\vfill
\leavevmode
\epsfysize=3cm
\epsffile{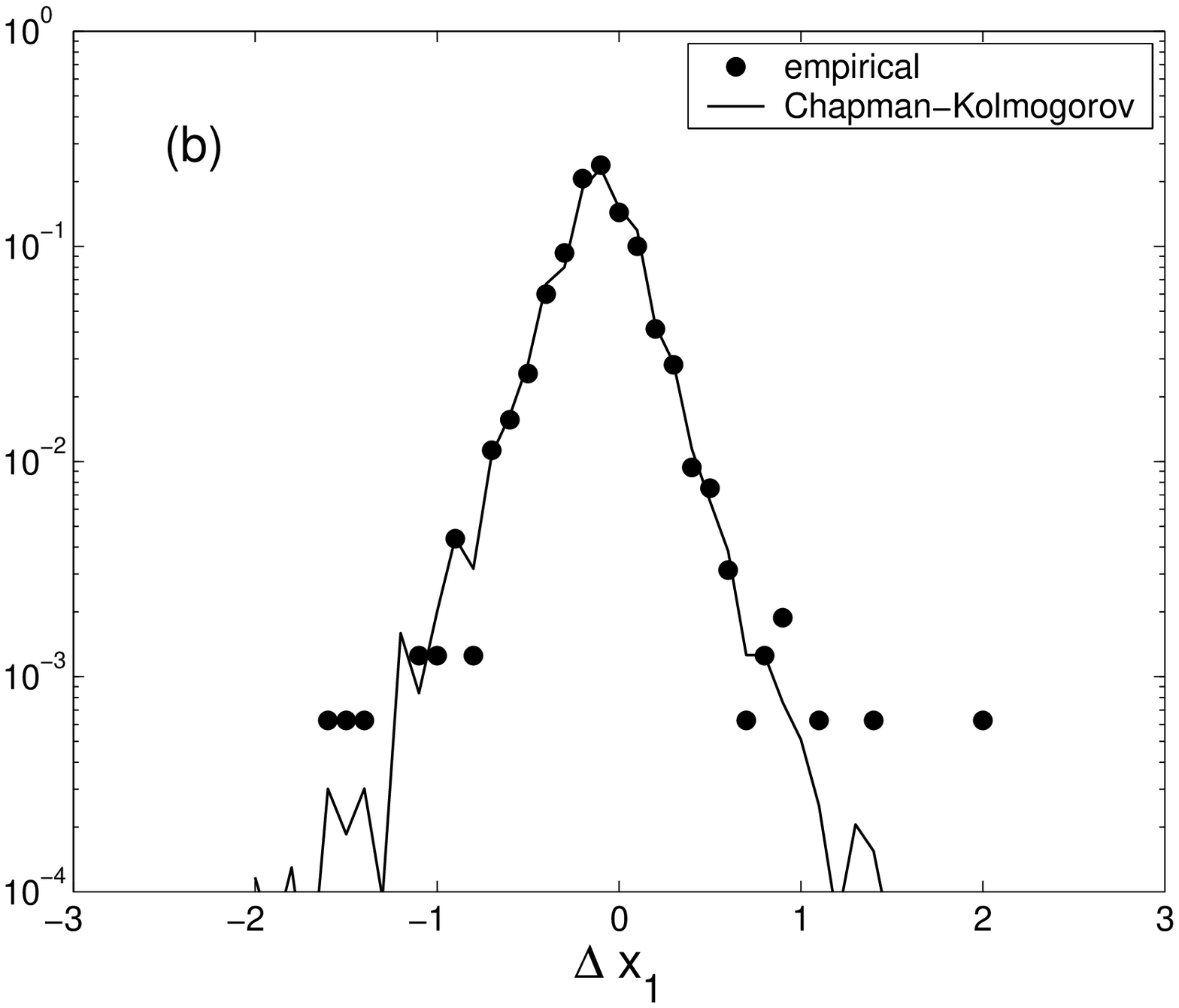}
\hfill
\leavevmode
\epsfysize=3cm
\epsffile{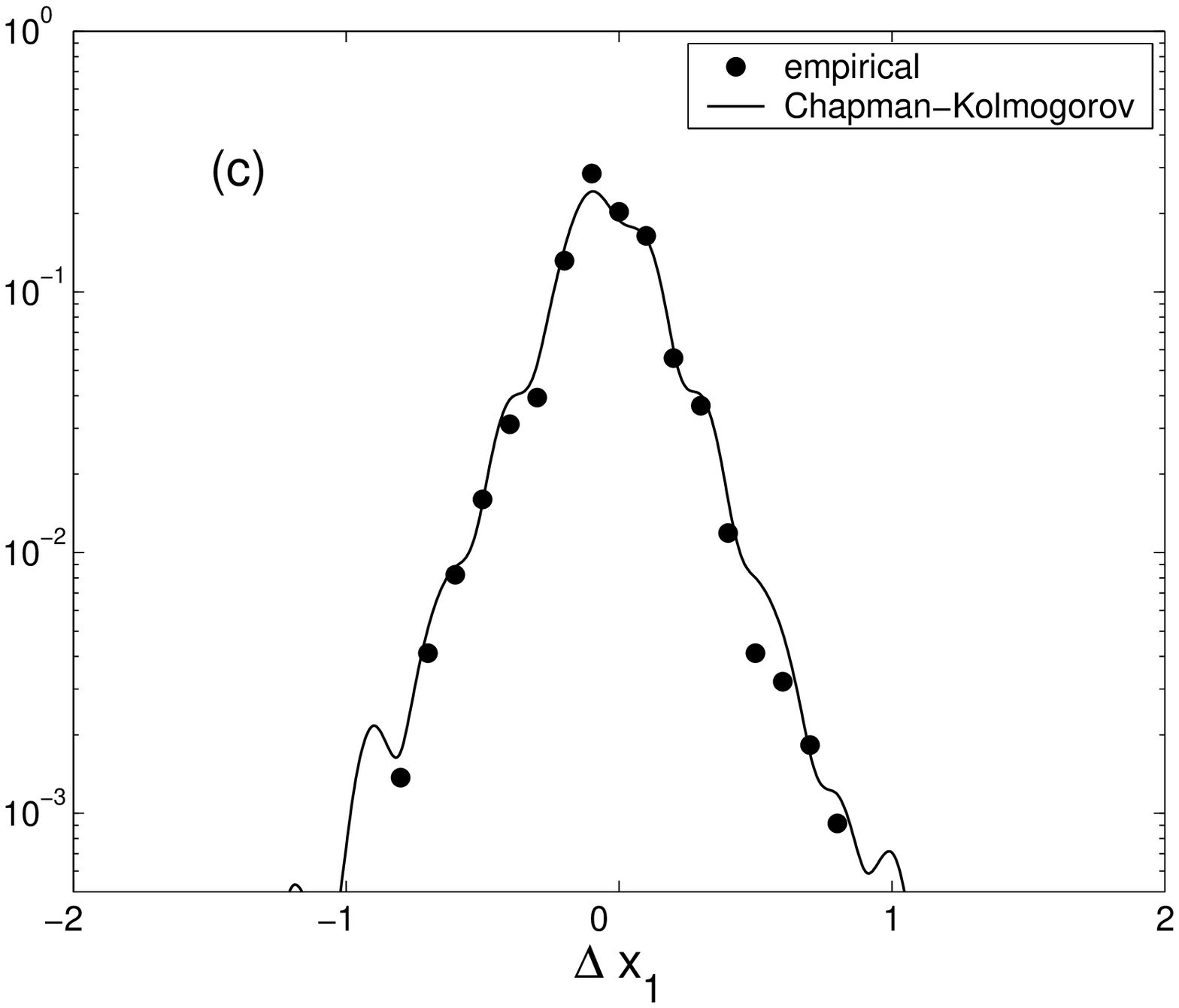}
\hfill
\leavevmode
\epsfysize=3cm
\epsffile{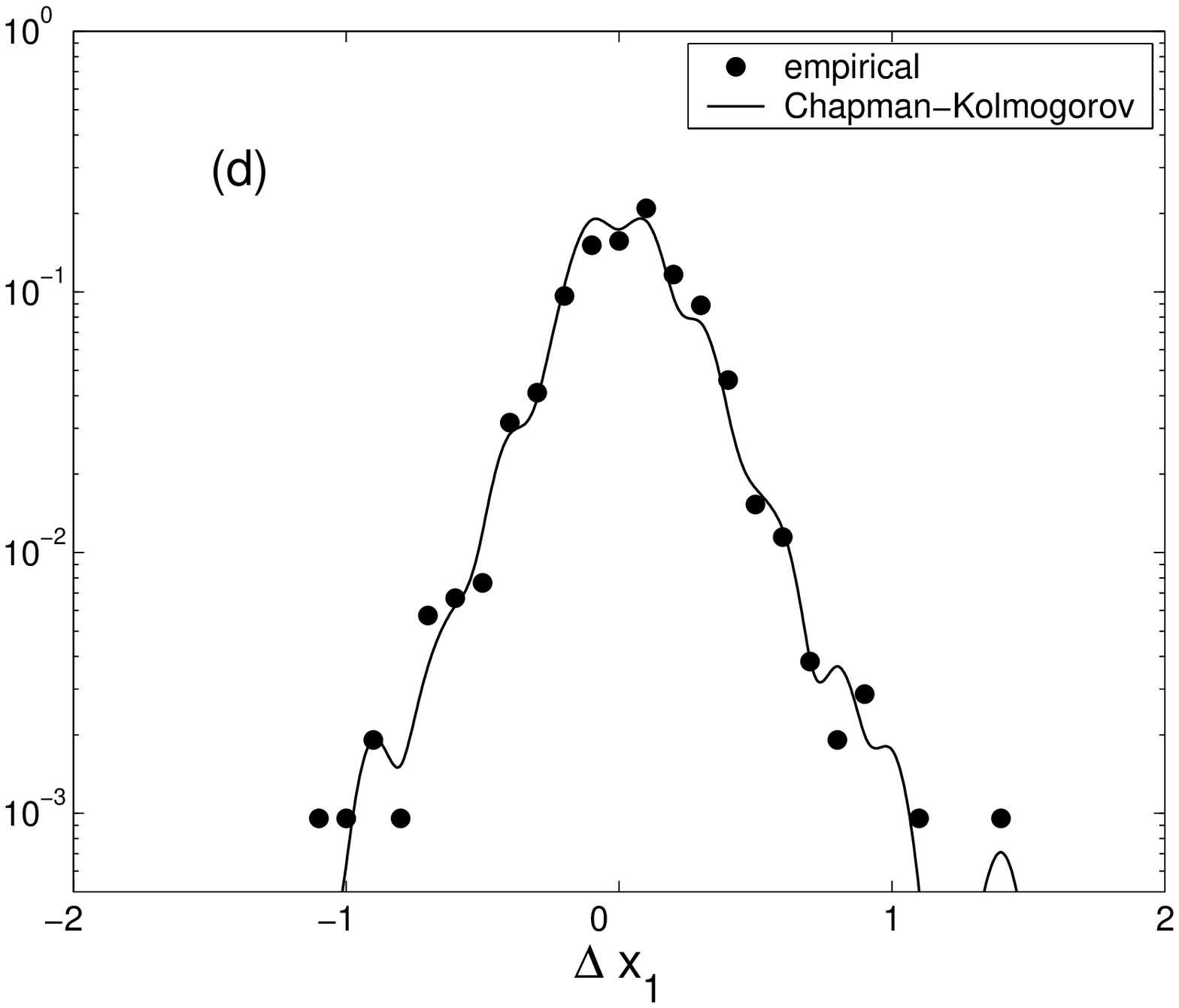}
\end{center}
\caption{(a) Empirical, and theoretical contour
plots of the conditional LWP increment pdf $p$($\Delta x_1$,$\Delta t_1$| $\Delta
x_2$,$\Delta t_2$) for $\Delta t_2$ = 180~s and $\Delta t_1$ = 20~s. In order to
verify the Chapman-Kolmogorov equation, the directly evaluated pdf (solid) is
compared with the integrated pdf (dotted). Assuming a statistical error of the
square root of the number of events of each bin, both pdfs are found to be
statistically identical; (b), (c), and (d) directly evaluated pdf's (dots) and
results of the numerical integration of the Chapman-Kolmogorov equation (line)
for cuts at $\Delta x_1$ = -0.4, 0.0, 0.4.}
\end{figure}

\begin{figure}[ht]
\begin{center}
\leavevmode
\epsfysize=8cm
\epsffile{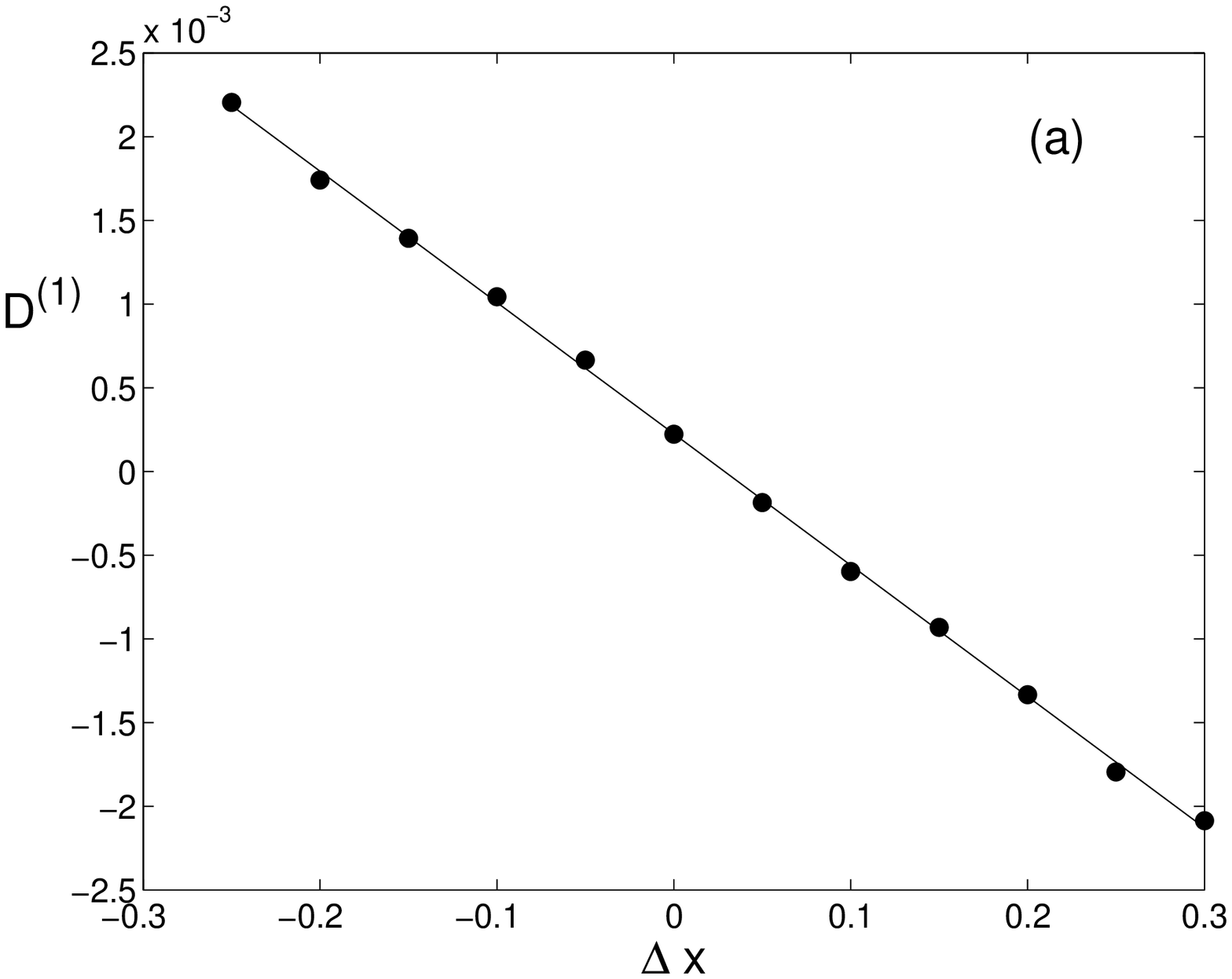}
\vfill
\leavevmode
\epsfysize=8cm
\epsffile{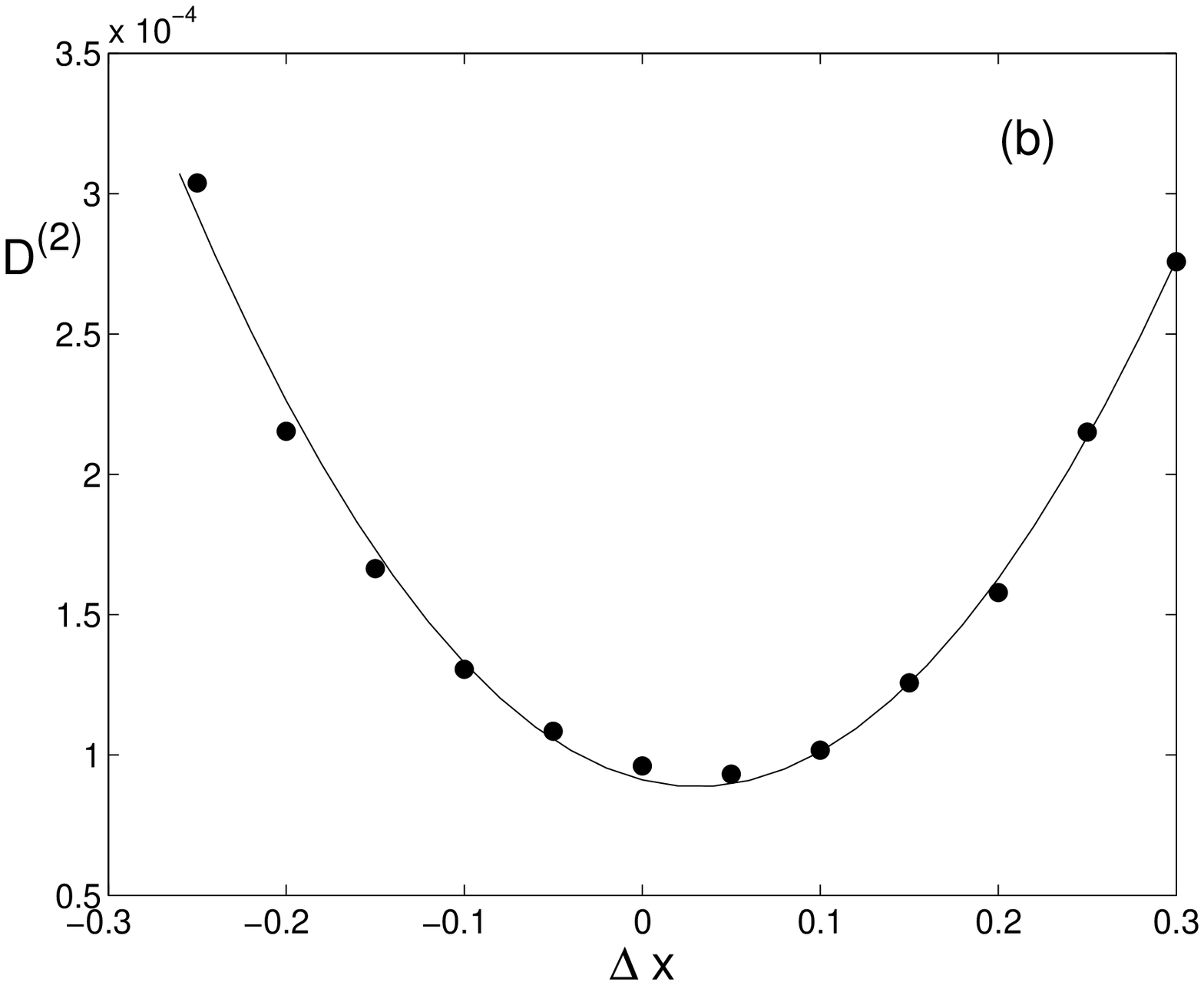} 
\end{center}
\caption{Kramers-Moyal coefficients (a)
$D^{(1)}$ and (b) $D^{(2)}$ estimated from conditional pdf $p(\Delta x_1,\Delta
t_1|\Delta x_2,\Delta t_2)$; $\Delta t_1=20$~s and $\Delta t_2=40$~s. The solid
curves present a linear and a quadratic fit, respectively.}
\end{figure}

\end{document}